\documentclass[nofootinbib,superscriptaddress,twocolumn,prl]{revtex4-1}

\usepackage{wrapfig}
\usepackage{graphicx}
\usepackage{float}
\usepackage{subfigure}
\usepackage{amssymb}
\usepackage{bbold}
\usepackage{color}
\usepackage{amsthm}
\usepackage{bm}
\usepackage{ulem}
\usepackage{comment}

\newcommand{\ket}[1]{\left|#1\right\rangle}

\newcommand{\id}{\mathbb{1}}

\usepackage{amsmath}

\usepackage[dvipsnames]{xcolor}
\usepackage[colorlinks=true,linkcolor=blue,citecolor=blue]{hyperref}

\makeatother

\begin{document}
\title{
Setting up experimental Bell test with reinforcement learning}

\author{Alexey A. Melnikov}
\affiliation{Department of Physics, University of Basel, Klingelbergstrasse 82, 4056 Basel, Switzerland}

\author{Pavel Sekatski}
\affiliation{Department of Physics, University of Basel, Klingelbergstrasse 82, 4056 Basel, Switzerland}

\author{Nicolas Sangouard}
\affiliation{Department of Physics, University of Basel, Klingelbergstrasse 82, 4056 Basel, Switzerland}
\affiliation{Institut de Physique Th\'eorique, Universit\'e Paris Saclay, CEA, CNRS, F-91191 Gif-sur-Yvette, France}

\begin{abstract}
Finding optical setups producing measurement results with a targeted probability distribution is hard as \textit{a priori} the number of possible experimental implementations grows exponentially with the number of modes and the number of devices. To tackle this complexity, we introduce a method combining reinforcement learning and simulated annealing enabling the automated design of optical experiments producing results with the desired probability distributions. We illustrate the relevance of our method by applying it to a probability distribution favouring high violations of the Bell-CHSH inequality. As a result, we propose new unintuitive experiments leading to higher Bell-CHSH inequality violations than the best currently known setups. Our method might positively impact the usefulness of photonic experiments for device-independent quantum information processing.
\end{abstract}

\maketitle

\paragraph{Introduction---} Bell nonlocality is a remarkable feature that allows to prove the incompatibility of an experiment with a classical description by only assuming no-signaling and some form of independence between the experiment's devices~\cite{Brunner14}. First appreciated for its foundational implications, it was recently realized that Bell nonlocality is also a valuable resource to perform some quantum information tasks in a device-independent way, i.e without ever assuming a detailed quantum model of the setup. This led to trustworthy protocols for quantum key distribution with device-independent security guarantees~\cite{Renner14} and random number generation with device-independent randomness certifications~\cite{Herrero-Collantes17}. In the same vein, self-testing~\cite{Supic2019} is the only technique, that allows to certify the functioning of quantum devices without assuming anything particular about their physical model. 

In device-independent protocols, the quantity of interest, be it the rate of quantum key distribution, the randomness of outcomes or the deviation from the ideal device in self-testing, is expressed as a scalar function of the probability distribution of measurement outcomes. In most cases, the targeted probability distribution is the one leading the highest Bell inequality violation. For example, the simplest and most studied Bell test is the Clauser, Horne, Shimony and Holt (CHSH) test~\cite{CHSH69} which can be used for all aforementioned tasks. From a practical perspective one of the main goals for performing device-independent quantum information tasks is thus to design an experiment with measurement outcomes optimizing the violation of the CHSH inequality.

With their high repetition rates and routinely controlled devices, optical experiments are appealing to perform a CHSH test as demonstrated in many
experiments~\cite{Giustina13,Christensen13,Shalm15,Giustina15,Shen18,Liu18}. The CHSH inequality violations reported in these experiments are very small which prevent one to use them for most applications of device-independent quantum information processing. It is thus natural to wonder if the same optical devices can be re-arranged to increase the CHSH inequality violations. Finding the setup leading to the highest CHSH score, that is the highest CHSH inequality violation, is however not straightforward, as too many possibilities must be considered. It is the case that automation is becoming necessary to improve photonic implementations of device-independent protocols.

Inspired by recent developments in machine learning~\cite{lecun2015deep,schmidhuber2015deep,human2015mnih,silver2018general} which is becoming more and more useful in automation of problem-solving in quantum physics research~\cite{dunjko2018machine,carleo2019machine,biamonte2017quantum,dunjko2020nonreview}, we introduce a technique based on the interplay between reinforcement learning~\cite{sutton2018reinforcement} and simulated annealing~\cite{van1987simulated} to design photonic setups maximizing a given function of probability distributions of measurement results. We illustrate the relevance of this method by applying it to probability distributions 
yielding high violations of the CHSH inequality with controlled resources. In essence, our algorithm not only re-discovered two known setups, but also discovered new, unexpected experimental settings leading to higher CHSH scores, and, to our knowledge, are not analogous to any known settings.\\

\begin{figure*}[ht!]
	\centering
	\includegraphics[width=1.\linewidth]{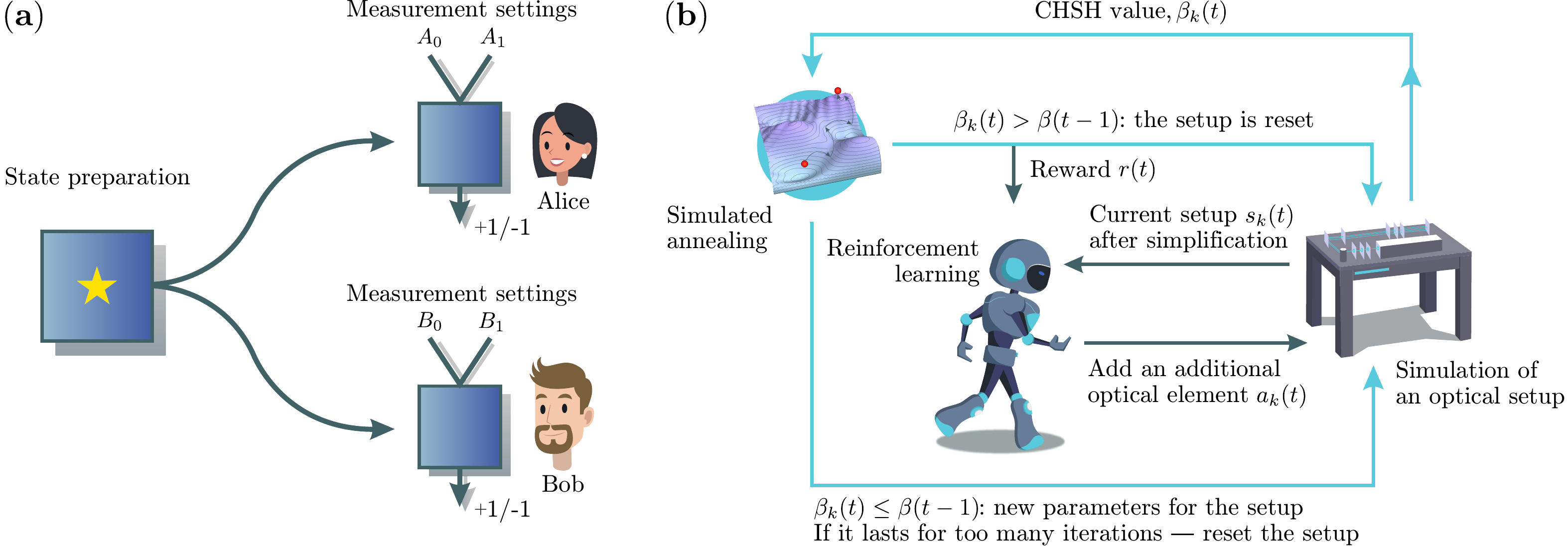}
	\caption{(a) Schematic representation of a Bell test with two parties -- Alice and Bob. In each round, Alice and Bob independently perform a measurement (chosen randomly among a two settings choice) on the state that they share. The measurement outcomes that Alice and Bob observe by repeating the measurements are used to test a Bell inequality. (b) Schematic representation of the proposed learning protocol to design photonic experiments leading to a probability distribution of measurement outcomes favouring a large CHSH inequality violation. Reinforcement learning (gray-green arrows) and simulated annealing (blue arrows) approaches are used together. Index $k$ corresponds to the agent-environment interaction step, whereas the variable $t$ -- to the trial. Once the setup is reset, and the new trial starts, only the agent's memory is preserved.
	}
	\label{fig:RLexp}
\end{figure*}

\paragraph{CHSH test---} The way to realize the CHSH test is depicted in Fig.~\ref{fig:RLexp}(a). A bipartite quantum state is prepared and shared between Alice and Bob. They then perform measurement on their system by randomly and independently choosing one out of two measurement settings: $A_x$ with $x=1,0$ for Alice and $B_y$ with $y=0,1$ for Bob. Each measurement has two possible outcomes $\pm 1$ labeled $a$ and $b$ for Alice and Bob, respectively. The CHSH score is then computed from the distribution of measurement outcomes as
\begin{equation}
\beta = \sum_{x,y=0}^{1} (-1)^{x y} \Big(p(a=b|A_xB_y)-p(a \neq b|A_xB_y)\Big)
\end{equation}
where $p(a=b|A_xB_y)$ is the probability that the results are the same given the setting choices $A_x$ and $B_y.$ Note that the CHSH score is upper bounded by 2 whenever there exists a locally causal model reproducing the measurement outcomes. The range of possible scores is enlarged to $|\beta|\leq 2\sqrt{2}$ with quantum models~\cite{Cirelson80}.\\

\paragraph{The physics of designed experiments---} We consider an experiment involving $n$ bosonic modes initialized in the vacuum state.  Their state is then manipulated by applying single-mode operations -- phase-shifters, displacement operations, and single-mode squeezers, and two-mode  operations -- beam-splitters and two-mode squeezers, on any mode or pair of modes in any order, or measuring out some of the modes with non photon-number resolving detectors, see Supplemental Material (SM), section A and B for a detailed description of these elements. The state preparation is complete if the desired combination of outcomes \textit{click} or \textit{no click} is observed on $n-m$ detectors. The remaining $m$ modes are shared between Alice and Bob, who locally apply a combination of operations from the same alphabet depending on their measurement settings. Finally, all the modes are measured out with non-photon number resolving detectors, yielding one of the $2^m$ possible outcomes. In our examples, we will consider the cases $\{m,n\}=\{2,2\}$ and $\{2,3\}.$ 

The alphabet of possible unitary operations we described is a fair representation of devices that are routinely used in optical experiments\footnote{The implementation of higher order interaction is way more demanding and  inefficient.}. It is worth mentioning that degrees of freedom, such as polarization or frequency are described by associating several bosonic modes to one photon. The use of single photon detector is motivated by the fact that the results of Gaussian operations alone can be reproduced by locally causal models.\\

\paragraph{Complexity---} To motivate an automated approach to our problem, let us discuss the complexity of finding the desired setup. A priori, the number of possible arrangements of elements grows exponentially with the number of modes and the total number of elements. However, the operations we consider belong to the class of Gaussian transformations, except the detectors. In consequence, any combination of such elements defines a Bogolyubov transformation on the $n$ modes. Reciprocally, any state prepared by the action of a Bogolyubov transformation on the vacuum can be prepared with $O(n^2)$ elements only, see SM section B for the details. Yet, an automated approach is very relevant for the design of such experiments for three reasons. First, the total number of parameters must also account for the measurement settings and a brute-force approach would have to optimize the parameters of 23 elements for finding the highest CHSH score for the case $\{m,n\}=\{2,3\}$. Second, if the elements include imperfections, the set of transformations which is accessible by combining individual elements is in general unknown, making a brute-force method unfeasible. Finally, a brute-force search is unsuitable when one is interested in keeping the number of elements low. \\

\begin{figure*}[ht!]
	\centering
	\includegraphics[width=0.95\linewidth]{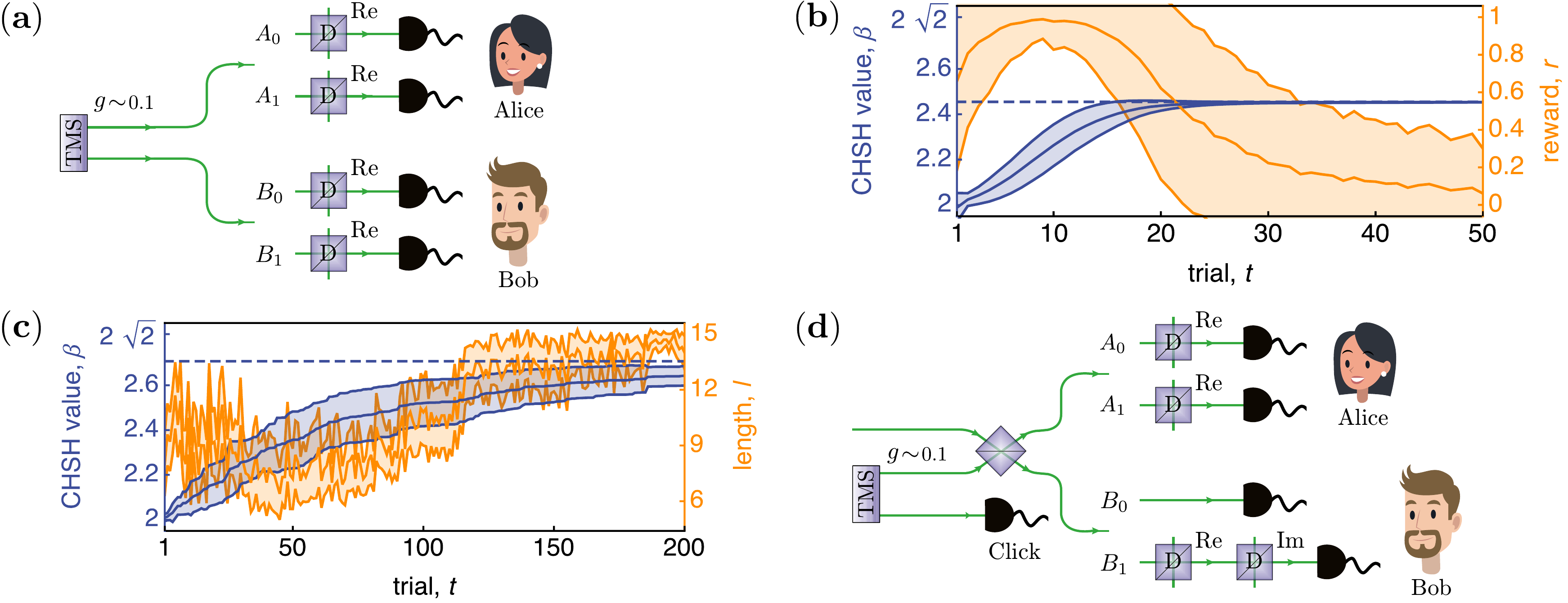}
	\caption{(a) Setup used to test the reliability of the simulated annealing based optimization. Two modes initially empty, are coupled by a two-mode squeezing (TMS) operation. One mode is sent to each party which he/she measures using photon detection preceded by displacement operations (D). (b) Learning curves reflecting the parameter optimization process performance for the fixed sequence shown in (a). The results are an average over $1000$ independent runs, shaded areas are mean squared deviations. $\beta\approx 2.4547$ (dashed line) is the optimal CHSH value found by using an analytical expression. (c) CHSH score and experiment length reflecting the reinforcement learning process of improving the CHSH value using a restricted set of elements. The results are an average over $10$ independent runs, shaded areas are mean squared deviations. The maximal value of $\beta\approx 2.7075$ found by the learning agent is shown as a dashed line. (d) Setup combining simplicity and a relatively high CHSH score $\beta\approx 2.6401$ (the heralding probability is $p_\mathrm{click}\approx 2.2\times 10^{-3}$). The diamond shape element above the heralding detector is a beam splitter.
	}
	\label{fig:SARLlearning}
\end{figure*}

\paragraph{Learning to violate the CHSH inequality---} We now formalize the problem of an automated design of an optical setup leading to a high CHSH score as a reinforcement learning~\cite{sutton2018reinforcement,melnikov2018active} task. The description of a setup we outlined before, possesses two levels. The top level specifies the order in which different elements are applied on different modes. As each element is parametrized by one or more parameters, a second level specifies the value of these parameters. The automated design we propose treats the two levels on a very different footing -- a learning agent focuses on the first level and an optimization algorithm based on simulated annealing treats the second level. As seen in Fig.~\ref{fig:RLexp}(b), the learning agent (shown as a robot) is interacting with a simulated optical experiment (shown as an optical table) in rounds, or interaction steps. A sequence of interaction steps that leads to a feedback signal (reward) is called a trial. One interaction step is visualized in Fig.~\ref{fig:RLexp}(b). At the beginning of each trial $t$ the agent perceives an input $s_1(t)$ (percept), which is a representation of an initial (empty) optical setup at the interaction step $k=1$. After a deliberation process, the agent chooses an optical element $a_1(t)$ (action) out of the set of available elements. The chosen element is incorporated into the setup $s_1(t)$ and this combination makes up a new setup $s_2(t)$. The latter is then analyzed with the help of an optimization technique in order to understand the quality of the prepared setup. The optimization technique based on simulated annealing~\cite{van1987simulated,nikolaev2010simulated,pham2012intelligent}, tries to find the best parameters of all the chosen optical elements in the setup such that the measurement results lead to the highest possible CHSH score. The search for parameters runs until the maximum number of optimization iterations is reached, or once the CHSH score at step $k$ is higher than the previous best one $\beta(t-1) = \max_{k',t'}\beta_{k'}(t'|t'\leq t-1)$. If $\beta_k(t) > \beta(t-1)$, a reward of $r(t)=1$ is given and the trial is finished: the agent starts the next trial from the initial configuration $s_1(t+1)$ with an updated CHSH score $\beta(t)$.\\

\paragraph{Optimizing a fixed setup---} The parameter optimization process is embedded into our reinforcement learning task. In particular, rewards -- the signals that the learning agent uses to improve its own performance in the design of experiments -- are the outputs of the parameter optimization procedure. The decision making relies on a good optimization process that provides near-optimal CHSH values for each setup proposed by the agent. We thus test the reliability of the simulated annealing based optimization by first separating the optimization process from the reinforcement learning process. This is implemented by considering a setup where Alice and Bob directly measure the state produced by vacuum two-mode squeezing using photon detection preceded by displacement operations in phase space, see Fig.~\ref{fig:SARLlearning}(a). The learning scenario is equivalent to having a ``lucky'' agent that always chooses a setup known to produce a high CHSH score~\cite{Kuzmich00,Lee09,Brask12}, or to an agent that already learned this configuration.

Fig.~\ref{fig:SARLlearning}(b) shows the results of the parameter optimization in this scenario. In our protocol the computation time allocated to the annealer increases with each subsequent trial. As a consequence, the CHSH score grows with the number of trials towards its maximum theoretical value (dashed line), which confirms that the optimization based on simulated annealing found near-optimal parameters. The maximum CHSH score found by the simulated annealing based optimization is $\beta\approx 2.4546$, which has only a $10^{-4}$ difference with the known optimal solution. Apart from the optimization procedure, it is important that a correct reward signal is paired with the dynamics of the CHSH score $\beta$. In Fig.~\ref{fig:SARLlearning}(b) one can see that reward $r$ is growing with time in parallel with $\beta$. However, the reward goes down towards zero with time, meaning that the agent receives nearly no feedback after $20$ trials. This unusual phenomenon, opposite to a standard scenario in reinforcement learning, where an agent is expected to keep maximizing the reward per trial, is explained by that fact that it becomes harder and harder to surpass the CHSH values of previous rounds once the CHSH score has nearly reached its maximum theoretical value. For the observed reason, we expect our learning agent to converge to an average reward of zero in the learning tasks that we consider next.\\

\paragraph{Reinforcement learning with a limited set of devices---} We first start by considering a task in which the agent is given $n=3$ bosonic modes and a restricted toolbox composed of beam splitters, two-mode squeezers and displacement operations. Given that $2$ modes will finally be distributed to Alice and Bob and the last one is detected, the agent is asked to find the setup leading to the highest CHSH score by getting feedback from the simulated annealing based optimizer.

As a reinforcement learning agent, we are using the projective simulation agent, which was first introduced in Ref.~\cite{briegel2012projective} and since then was shown to be attractive both theoretically~\cite{mautner2013projective,makmal2016meta,melnikov2017projective,clausen2019convergence} and for practical use~\cite{hangl2016robotic,melnikov2018active,melnikov2018benchmarking,wallnofer2019machine,nautrup2019optimizing}. The details of the agent's internal structure and meta-parameters are omitted here, see SM section A for more information on the implementation. The code of the basic projective simulation agent is publicly available in Ref.~\cite{PScode}. 

Fig.~\ref{fig:SARLlearning}(c) shows the evolution of the CHSH score and the length of the experiment $l(t),$ i.e. the evolution of the number of elements used by the agent per trial. One can see that the setup length first decreases before increasing. Similar to the behavior of the reward $r(t)$ observed in Fig.~\ref{fig:SARLlearning}(b), it is hard to improve over nearly optimal solutions, hence the agent gets no reward and by gradually forgetting its previous experience, explores the most complex possibilities of the maximum allowed length $l_\mathrm{max}=15$. This maximum length is the same throughout our work, and is lower than the number of $l=23$ independent elements required for a general Bogolyubov transformation on the 3 modes, see SM section B for the  details. As for the CHSH values, a setup is found leading to a maximum value of $\sim 2.7075$ using eleven elements. Although, the agent gets rewards more frequently when the CHSH score tends to increase which may favor long lengths, he effectively gets a similar reward per action in shorter experiments with lower CHSH values. This trade-off can be controlled by the agent's meta-parameters. When short length setup are favored, the agent  finds frequently a simple setup leading to a CHSH score of $\sim 2.6401$ in which the modes 2 and 3 are first coupled with a two mode squeezer and the modes 1 and 2 are then coupled through a beam splitter, see Fig.~\ref{fig:SARLlearning}(d). This corresponds to a physical implementation of a setup proposed in 1999 ~\cite{Banaszek99} where a CHSH inequality violation is observed by first sending a single photon into a beam splitter and measuring the output modes with photon detection preceded by displacement operations. Note that the agent is using displacements with real and imaginary parts, however, does not use them symmetrically in the measurement choices. Each element with a complex parameter is indeed divided into two elements, one with a real parameter and one with a purely imaginary parameter and the agent found a way to reduce the number of elements in the displacements from $8$ to $4$. This represents the internal feature of reinforcement learning agents -- reducing the number of actions per reward, which corresponds to reducing the number of elements in setups. \\

\begin{figure}[ht!]
	\centering
	\includegraphics[width=1\linewidth]{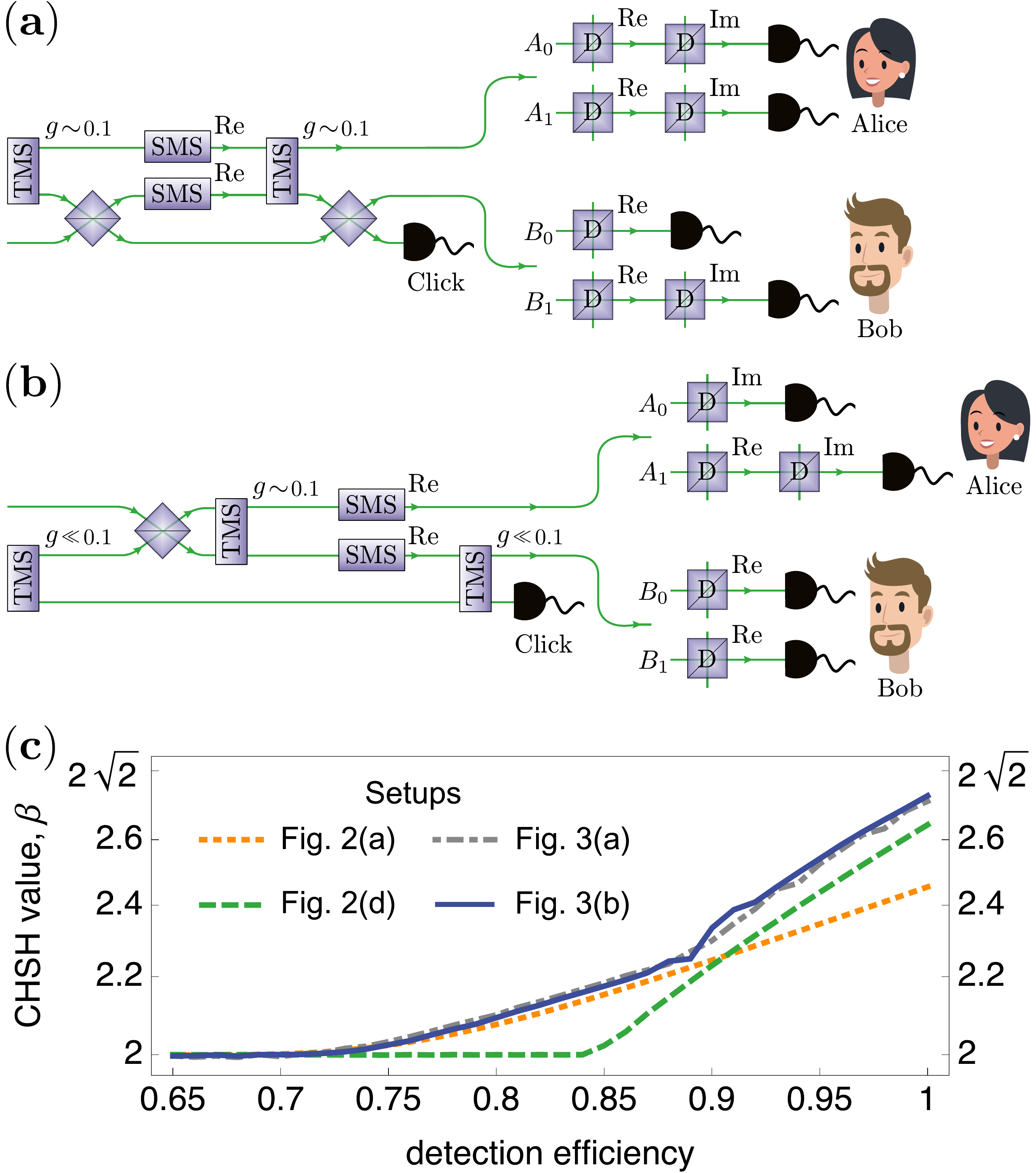}
	\caption{Two photonic setups designed by the projective simulation agent after learning from simpler examples. SMS elements correspond to single-mode squeezing operations. (a) A Bell test with CHSH score $\beta\approx 2.7242$ and a heralding probability $p_\mathrm{click}\approx 2.9\times 10^{-4}$, or $\beta\approx 2.7424$ with $p_\mathrm{click}\approx 2.6\times 10^{-5}$. (b) A Bell test with $\beta\approx 2.7454$ and $p_\mathrm{click}\approx 1.1\times 10^{-9}$. Parameters of the elements for both setups are in SM section A. (c) Dependence of the CHSH scores of setups which are considered in this work, on detection efficiency. The setups designed by the agent provide higher CHSH values than known setups for any detection efficiency.}
	\label{fig:NovelCHSHSetups}
\end{figure}

\paragraph{Proposal of new experiments with reinforcement learning---}
Next, we allow the agent to choose between more elements by adding single-mode squeezing with a complex squeezing parameter to the previous set of elements. 
Benefiting from the extended space of possibilities, the agent gives us new setups, two of them producing CHSH scores above $2.74,$ see Fig.~\ref{fig:NovelCHSHSetups}(a) and Fig.~\ref{fig:NovelCHSHSetups}(b). We were not able to find known setups producing similar states. In order to witness the relevance of these new setups, we computed their CHSH score for non-unit detection efficiencies and re-optimized systematically the parameters of each element when the detection efficiency is changed. For comparison, we also reported in Fig.~\ref{fig:NovelCHSHSetups}(c) the CHSH score of setups shown in  Fig.~\ref{fig:SARLlearning}(a) and Fig.~\ref{fig:SARLlearning}(d) which are known to be resistant to detector inefficiency and to produce high CHSH scores for close to unit efficiency detection respectively~\cite{Caprara15}. We conclude that the new setups provide higher CHSH scores for any detection efficiency. This can be partially understood by noting that the new setups have more elements and the possibility to control their parameters allows one to reduce them to setups close to the ones of Figs.~\ref{fig:SARLlearning} (a) and (d). Although more detailed analysis are needed to conclude about the usefulness of these new setups in practice, they might provide additional motivations to develop programmable photonic integrated circuits. \\

\paragraph{Conclusion and outlook---} We have introduced a new approach to design photonic quantum experiments maximizing a desired function of the probability distribution of measurement outcomes. Our approach combines the reinforcement learning technique of projective simulation and the optimization algorithm of simulated annealing which makes it possible to simultaneously explore and optimize the discrete space of optical elements and the continuous space of their parameters. The relevance of this machine learning technique has been shown by focusing on the design of experiments favoring high CHSH scores. We observed that the agent is able to learn the experimental designs by trial-and-error. The agent was designing setups with increasing CHSH value and decreasing experiment length whenever it was possible. As a result, new setups have been discovered with unprecedented CHSH values for any detection efficiency, which might positively impact the usefulness of photonic experiments for device-independent quantum information processing. \\

\paragraph{Acknowledgment---}
We are thankful to Enky Oudot, Xavier Valcarce, and Jean-Daniel Bancal for useful discussions. This work was supported by the Swiss National Science Foundation (SNSF), through the Grant PP00P2-179109 and by the Army Research Laboratory Center for Distributed Quantum Information via the project SciNet. Computation was performed at sciCORE (scicore.unibas.ch) scientific computing core facility at University of Basel.


%

\section{Supplemental Material}\label{SM}

\subsection{A-Details on the learning algorithm}
This part of the Supplemental Material is divided into 4 paragraphs. The first one gives the details on the optical setups simulation, the second one focuses on the simulated annealing algorithm, the third one is dedicated to the projective simulation model. The values of parameters that were obtained for the setups presented in the main text are given in the last paragraph. \\

\begin{figure*}[ht!]
	\centering
	\includegraphics[width=0.9\linewidth]{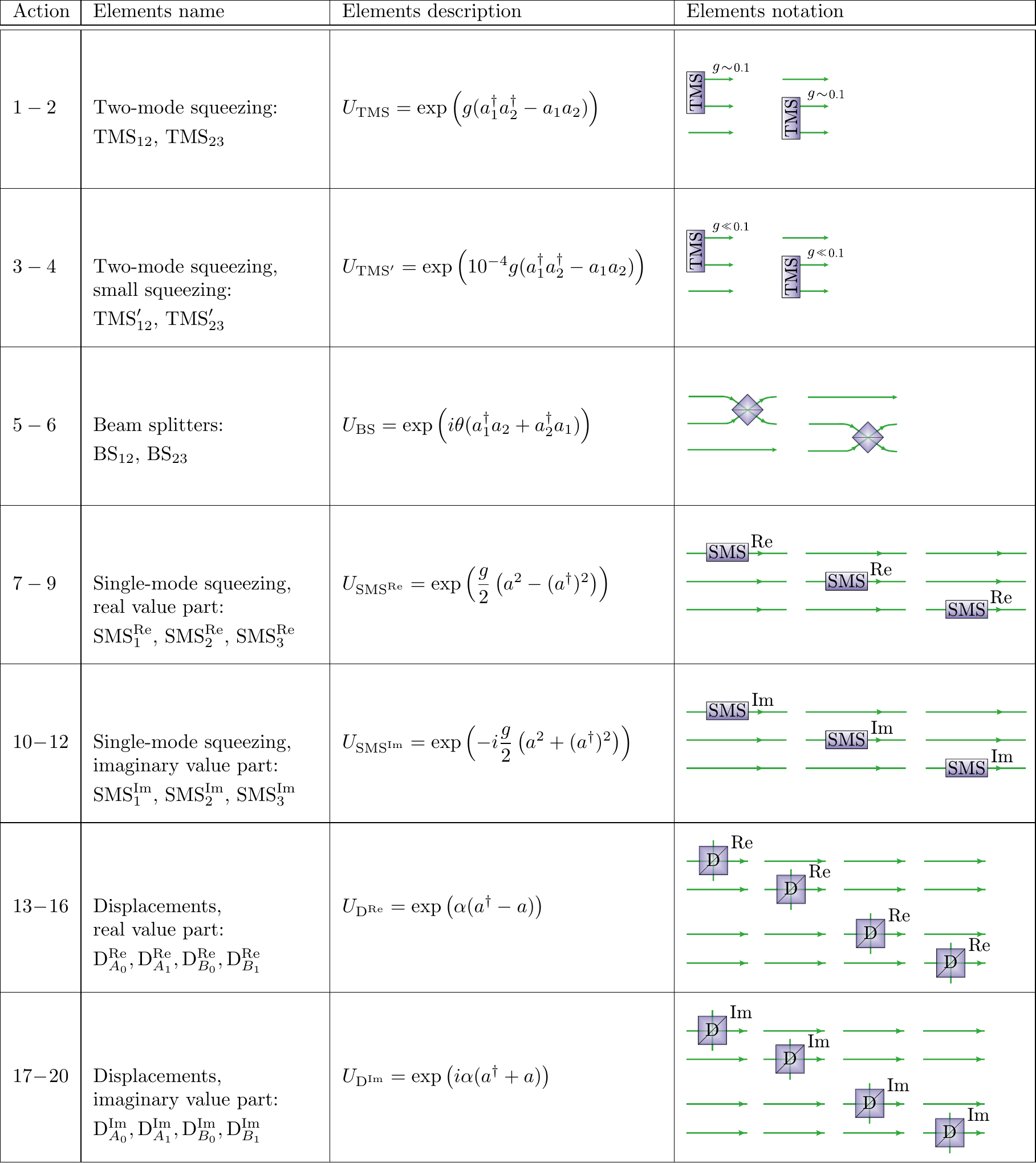}
	\caption{Optical elements that are used by the projective simulation agent as $20$ different actions. All parameters in the unitary operations are set by the simulated annealing algorithm.}
	\label{fig:ActionsTable}
\end{figure*}

\paragraph{Optical setups simulation---} In this section we give details on the way the simulation of optical setups is performed. This part of our work is schematically shown as an optical table in Fig.~\ref{fig:RLexp}(b) and involves all arrows going in and out of this table. The gray-green arrows in Fig.~\ref{fig:RLexp}(b) correspond to the interaction of the reinforcement learning agent with the simulation. This interaction at step $k$ involves giving an optical element $a_k$ to the simulator, and getting a setup $s_k$, which is a simplified sequence of previously set elements. All the actions $a$ that are in the agent's toolbox $\mathcal{A}$ are shown in Fig.~\ref{fig:ActionsTable}. There are in total of $|\mathcal{A}| = 20$ actions, which correspond to specific optical elements shown in the right column. These optical elements also represent unitary operations, written in the third column. Each of the unitary operations has a single parameter: $g$, $\theta$, or $\alpha$, and can take any real value. Practically, the values of the parameters are bounded by the number of the simulated annealing based optimization steps: at each step, a parameter can be changed by maximal value of $0.25$. Unitary operations act on the state of photonic modes that are modeled as vectors in the $d$-dimensional Hilbert space, which means we consider up to $d-1$ photons per mode. $d$ is set to $4$ in all our simulations, however, in the event of non-zero reward, we redo the simulations automatically with $d=11$.

The described actions are arranged in a sequence $\{a_1(t), a_2(t), \dots, a_k(t)\}$, forming a photonic setup $s_k(t)$. From this sequence we delete all the duplicates of the following displacement elements: $\mathrm{D}^\mathrm{Re}_{A_0}, \mathrm{D}^\mathrm{Re}_{A_1}, \mathrm{D}^\mathrm{Re}_{B_0}, \mathrm{D}^\mathrm{Re}_{B_1}, \mathrm{D}^\mathrm{Im}_{A_0}, \mathrm{D}^\mathrm{Im}_{A_1}, \mathrm{D}^\mathrm{Im}_{B_0},$ and $\mathrm{D}^\mathrm{Im}_{B_1}$. This duplicates deletion operation simplifies the percept representation $s_k(t)$, and does not introduce any change to the setup as all possible unique experiments can be achieved by one copy of the mentioned displacement operations. Sometimes, however, the agent decides to add more displacements of the same type in order to improve the precision of the optimization. Placing the same element twice does not change the percept, but makes another optimization run which gives a higher chance to find a better CHSH value. We have observed that this effect appears once the learning agent learns a near-optimal setup.

To avoid computational complexity in simulations, and to avoid over-complicated photonic setups that are difficult to realize experimentally, we set the maximum length of a setup to $l_\mathrm{max}=15$ and the maximum number of interaction steps within a trial to $k_\mathrm{max}=20$. If, after creating a setup with $15$ elements, the learning agent does not get a positive reward, the constructed setup is reset. The same happens if the agent applies $20$ actions in one trial. 

In all the experiments that the learning agent designs, the third optical mode is measured with a non-discriminating photon detector at the state preparation stage. This heralding detector is the same as the detectors on Alice and Bob sides: the outcome is \textit{click} in case one or more photons are detected, or \textit{no click} otherwise, with POVM elements
\begin{align}
    P_\mathrm{click} &= \sum_{k=1}^\infty |k\rangle \!\langle k|\\
    P_\mathrm{no~click} &= |0\rangle \!\langle 0|.
\end{align}
For two measurement outcomes of the heralding detector both quantum states are considered and two CHSH values are computed. A measurement outcome corresponding to a quantum state with the highest CHSH value is kept, together with the corresponding probability $p_\mathrm{click}$, or $p_\mathrm{no~click}$. It is worth noting that discarding the outcome of the heralding detector -- or simply removing it -- can only be detrimental for the CHSH value, as follows from linearity of quantum mechanics.
It is the case, especially when single-photon experiments are involved, that the detector ``clicks'' with a very small probability below $10^{-10}$. The agent actively exploits this regime by learning to find setups where computational noise perturbs detection probabilities, leading to nonphysical quantum states which give high CHSH score. In order to avoid negative side effects with numerical precision, and without increasing the memory size for quantum states representation, we set a limit on the detection probability: $p^\mathrm{min}_\mathrm{click}= 10^{-4}$ in all the cases, except the one in Fig.~\ref{fig:NovelCHSHSetups} where $p^\mathrm{min}_\mathrm{click} = 10^{-9}$.

In all the learning settings the agent designs photonic setups for unit detector efficiencies and zero photon loss. In Fig.~\ref{fig:NovelCHSHSetups}(c) we, however, additionally study the effectiveness of the designed setups for detection efficiency as modeled by POVM elements
\begin{align}
    P_\mathrm{click}&= 1 - P_\mathrm{no~click} \\
    P_\mathrm{no~click} &= (1-\eta)^{a^\dag a}.
\end{align}
In the latter case, each photon arrives at a detector with probability $\eta$, which corresponds to the detector efficiency in Fig.~\ref{fig:NovelCHSHSetups}(c). \\

\paragraph{Simulated annealing algorithm---}
In this section we give details on the way the simulated annealing algorithm is implemented in our work.
The purpose of this algorithm is to set the parameters for a given photonic setup such that the CHSH score $\beta$ is maximized. Optimization algorithms are known to be useful not only for setup parameter optimization, but also for full quantum experiment design~\cite{krenn2016automated,knott2016search,driscoll2019quantum,nichols2019designing,krenn2020computer}. In our learning approach, different to previous works, optimization takes a role of a tool for the reinforcement learning agent to design better experimental setups. 

Simulated annealing algorithm part of our work is schematically shown as a cost function landscape in Fig.~\ref{fig:RLexp}(b) and involves all arrows going in and out of this function landscape. At a given interaction step $k$ of a trial $t$, the simulated annealing based algorithm is optimizing $\beta$ over the parameters of the setup $s_k(t)$. The photonic setup is simplified, therefore the length of the setup $s_k(t)$ is $l_k(t)\leq k$. Hence, simulated annealing is optimizing parameters $\Phi(k,t) = \{\phi_1(k,t), \phi_2(k,t), \dots, \phi_l(k,t)\}$ of the setup $s_k(t)$ in an $l$-dimensional parameter space. 

\begin{figure}[ht!]
	\centering
	\includegraphics[width=1\linewidth]{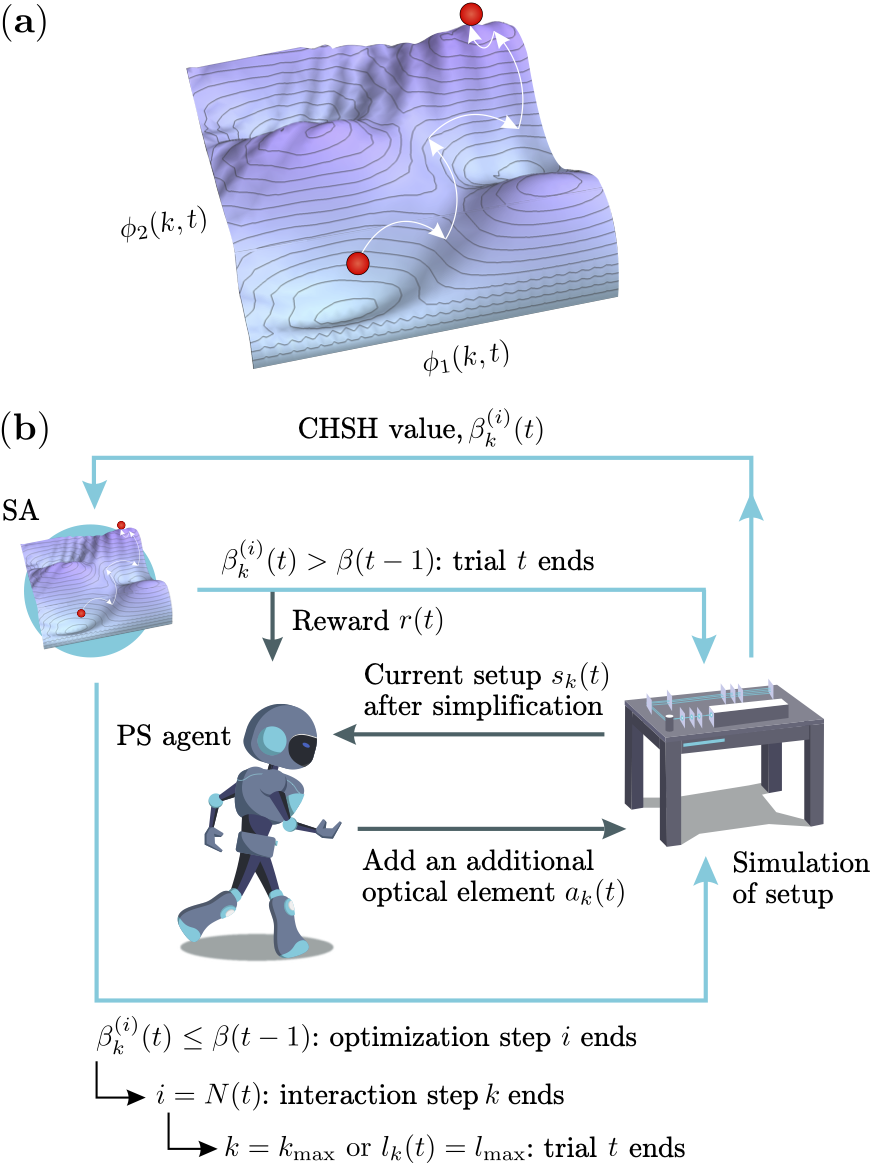}
	\caption{(a) Simulated annealing algorithm visualization for two-parameter optimization. Height corresponds to the CHSH value $\beta$. (b) A scheme that summarizes additional details of the machine learning approach that is used throughout the paper. In this scheme the emphasis is made on the parts which are controlled by the SA algorithm. Index $i$ corresponds to the optimization step (SA arrows), index $k$ -- to the agent-environment interaction step (RL arrows), and variable $t$ -- to the trial. Once the setup is reset, and the new trial starts, only the RL agent's memory is preserved.}
	\label{fig:SAalgorithm}
\end{figure}

The optimization starts at the point where all parameters are equal to zero, and continues in iterations. Iteration $i$ is implemented in the following way. First, one of the parameters $\phi_{m(i)}(k,t)$ is chosen uniformly from the parameter set $\Phi(k,t)$, and $\phi_{m(i)}(k,t)$ is changed to $\phi_{m(i)}(k,t)+\xi_i$; $\xi_i\in \mathbb{R}$ is sampled from the interval of $[-0.25,0.25]$. Next, $\beta^{(i)}_k$ value is calculated by the optical setups simulator. If $\beta^{(i)}_k(t)$ is larger than the highest CHSH score $\beta_k(t)$ during the step $k$, then the iteration ends and $\beta_k(t) = \beta^{(i)}_k(t)$ is updated\footnote{Initially, at the first iteration, $\beta_k(t)$ is set to zero.}. In case $\beta^{(i)}_k(t) < \beta_k(t)$, the change in the parameter $\phi_{m(i)}(k,t)$ is reverted with probability 
\begin{equation}
    p^{(i)}_k(t) = 1-\mathrm{exp}\left( -\frac{\beta_k(t) - \beta^{(i)}_k(t)}{T_i(t)} \right),
\end{equation}
where 
\begin{equation}
    T_i(t)=\frac{N(t)}{i}T_\mathrm{min}
\end{equation}
is the effective temperature of simulated annealing with $T_\mathrm{min}=0.001$ being the minimal temperature. $N(t)=95+5t$ is the total maximal number of optimization iterations done at trial $t$. After this probabilistic reversion of the parameter, the iteration $i$ ends. 

All iterations are continued up to a total number of iterations $N(t)$, or until the CHSH score is larger than the previous best value $\beta_k(t) > \beta(t-1)$. 
In case $\beta_k(t) > \beta(t-1)$, the trial $t$ ends, the setup is reset to $s_0(t+1)$, and a reward $r(t)$ is given to the agent. If the optimization stops after $N(t)$ iterations, and without achieving a better $\beta$, than the agent-environment interaction step $k$ finishes. But the trial $t$ continues by asking the agent to add one more element with an action $a_{k+1}(t)$. However, in case it was already the last interaction step (i.e. $k=k_\mathrm{max}$) or the length of the setup was already maximal (i.e $l_k(t)=l_\mathrm{max}$), the trial $t$ finishes with no reward and the setup is reset.

A visualization of a $2$-dimensional parameter space, as an example, is shown in Fig.~\ref{fig:SAalgorithm}(a). The particle (red) starts close to the local minimum of the CHSH value, followed be a jump along the $\phi_1(k,t)$ axis. This jump corresponds to the first parameter change. Then, the jumps are further performed along the axes until $N=5$ optimization steps. A summary of all the learning steps that the simulated annealing optimization algorithm influences is shown in Fig.~\ref{fig:SAalgorithm}(b).

\begin{figure*}[ht!]
	\centering
	\includegraphics[width=1\linewidth]{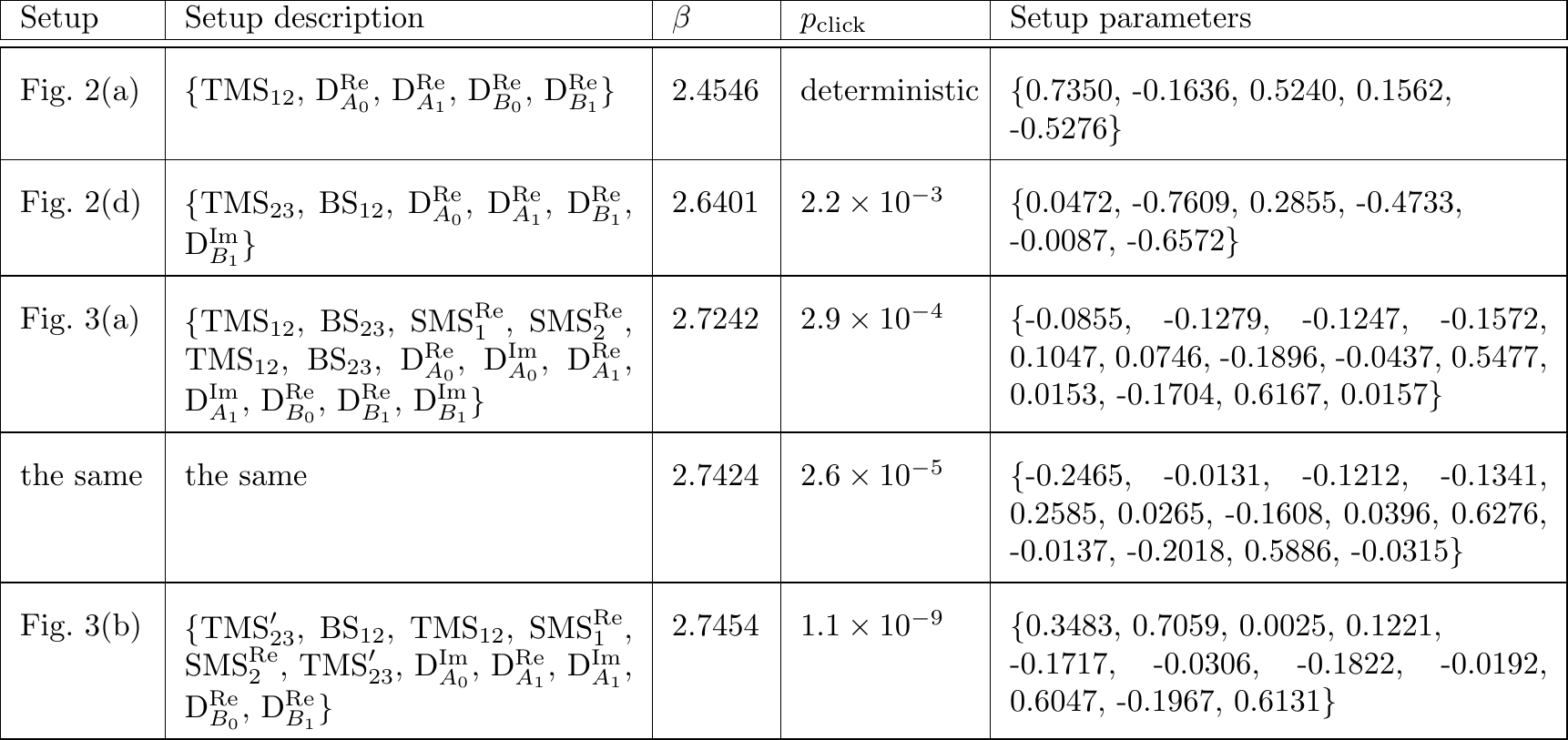}
	\caption{Four photonic setups that are discussed in the paper. One of the setups has two different parameter sets, which were obtained by setting a different constraint on the minimum detection probability $p_\mathrm{click}$.
	}
	\label{fig:SetupParameters}
\end{figure*}

Note that the parameters are changed by a value of $\xi_i\in [-0.25,0.25]$, which makes it practically difficult to explore the region of $\phi_{m(i)}(k,t)\ll 1$, or $\phi_{m(i)}(k,t)\gg 1$. This difficulty is naturally avoided by adding actions for different parameter scale. The actions $\mathrm{TMS}_{12}$ and $\mathrm{TMS}'_{12}$, $\mathrm{TMS}_{23}$ and $\mathrm{TMS}'_{23}$ (see table in Fig.~\ref{fig:SetupParameters} for the meaning of these acronyms) differ only in the scale of the squeezing parameter: in the case of $\mathrm{TMS}'$ squeezing parameter is on average $10^4$ times smaller.\\

\paragraph{Projective simulation model---} In this section we give details on the way our reinforcement learning algorithm is implemented. Reinforcement learning was shown to be able to design quantum experiments in Ref.~\cite{melnikov2018active}, where an agent was choosing from a discrete set of optical elements, but did not use any additional optimization technique to deal with continuous space of optical elements parameters.

Reinforcement learning part of our work is schematically shown as a robot in Fig.~\ref{fig:RLexp}(b) and Fig.~\ref{fig:SAalgorithm}(b). From the mathematical point of view, the learning agent can be viewed as a policy function $\pi_k(s,a,t)$ that maps input states $s_k(t)$ to output actions $a_k(t)$. Here we discuss the specific form of this policy functions, and it's updating.

\begin{figure}[ht!]
	\centering
	\includegraphics[width=1\linewidth]{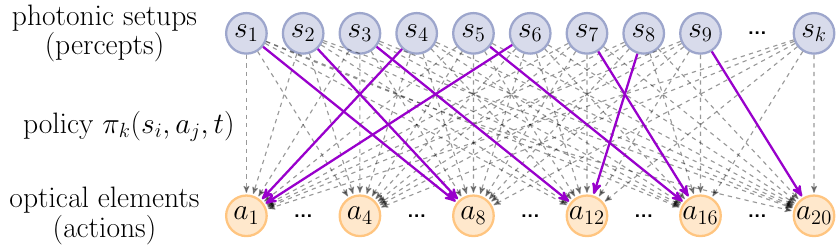}
	\caption{The PS network that is used in the paper. The network consists of two layers of clips and edges connecting these layers.}
	\label{fig:RLalgorithm}
\end{figure}

As a reinforcement learning agent, we use the projective simulation agent~\cite{briegel2012projective,mautner2013projective,makmal2016meta,melnikov2017projective,melnikov2018benchmarking,clausen2019convergence,hangl2016robotic} which is schematically represented in Fig.~\ref{fig:RLalgorithm}. The agent is a two-layered projective simulation network, similar to the one used in Refs.~\cite{melnikov2018active,wallnofer2019machine} for designing quantum experiments and quantum communication networks. The network consists of a layer of clips that represent states $s_k(t)$, and a layer of clips that represent actions $a_k(t)$. Clips are connected by directed edges with time-dependent weights $h_k(s,a,t)$. The weights fully define the policy function. The mapping from the weights $h$ to the policy $\pi$ can be done in various ways within the projective simulation model. In this work, it is done in the following way:
\begin{equation}
    \pi_k(s,a,t) = \frac{\mathrm{e}^{h_k(s,a,t)}}{\sum_{a'}\mathrm{e}^{h_k(s,a',t)}}.
\end{equation}
The policy is updated after each agent-environment interaction step $k$ by performing a change of weights $h_k(s,a,t)$:
\begin{eqnarray}
h_{k+1}(s,a,t) = h_k(s,a,t) - \gamma_\mathrm{PS}(h_k(s,a,t)-1) \nonumber
\\
+ g_{k+1}(s,a,t)r(t),
\label{updaterule}
\end{eqnarray}
where initial weights are set to $h_{1}(s,a,1)=1$ for all $(s,a)$ pairs. Reward $r(t)$ has a binary value: $r(t)=1$ in case the agent found a setup observing a better $\beta$ value during trial $t$, and zero otherwise. The values $g_k(s,a,t)$, which are used in the update of the weights, are called glow values. The glow value is set to $g_{k+1}(s,a,t)=1$ in case the state $s$ and action $a$ appeared in the step $k$. For all the other state-action pairs, the glow values are set to $g_{k+1}(s,a,t)=(1-\eta_\mathrm{PS})g_k(s,a,t)$ with $\eta_\mathrm{PS} = 0.3$. Initially all glow values are equal to zero as there is no history of state-action pairs.

The second term on the right-hand side of Eq.~\ref{updaterule} is a weight damping term that is responsible for forgetting. In our reinforcement learning scheme agent's forgetting helps to avoid locally (in the space of photonic setups) optimal CHSH values. Small level of forgetting $\gamma_\mathrm{PS}=10^{-3}$, which is the same in all our simulations, allows the agent to switch to completely new strategies in designing experiments within one learning run. Both $\gamma_\mathrm{PS}$ and $\eta_\mathrm{PS}$ are the meta-parameters of the projective simulation model. Meta-parameters are set to constant values in this work, but can also take trial-dependent values and be set by the agent autonomously~\cite{makmal2016meta}.\\

\paragraph{Details on learned setup parameters---} The values of the parameters that were obtained for the setups that are presented in the main text are detailed in the table given in Fig.~\ref{fig:SetupParameters}.

\subsection{B-Details on the problem complexity}

Take our $n$ bosonic modes with associated creation operators $a_1$ to $a_n$, satisfying the canonical commutation relations $[a_k,a^\dag_\ell]=\delta_{kl}$. For compactness, we combine the operators in a column vector denoted ${\bf a} = (a_1,\, a_2, \, \dots a_n)$. Pure states of the $n$ modes are elements of the Hilbert space $\mathcal{H}=F^{\otimes n}$ -- the tensor product of $n$ Fock spaces. Let us first look at different unitary operations from agent's alphabet in Heisenberg's picture through their action on $\bf a$. We start by linear optical transformations and then consider displacement and squeezing operations. We conclude on the elements that are required for implementing a general Bogolyubov transformation and end up with a precise count of the total number of parameters needed to describe all inequivalent experiments with a fixed number of measurement settings.  In particular, this gives  the total number of parameters that needs to be optimized in a brute-force search of the setup leading to the highest CHSH score.\\

\paragraph{Decomposition of an arbitrary linear transformation---} Let us start with phase shifters. A phase shifter -- denoted $\text{PS}$ -- is a single-mode single parameter $(\varphi)$ operation mapping a bosonic operator $a$ into
\begin{equation}
    \text{PS}_\varphi[a] = e^{ i  \varphi} a.
\end{equation}
A beam splitter -- labelled $\text{BS}$ -- is a two-mode single parameter $(\theta)$ operation mapping bosonic modes $a_1$ and $a_2$ into
\begin{equation}
    \text{BS}_\theta\left[\binom{a_1}{a_2}\right] = \binom{\cos(\theta) a_1-\sin(\theta) a_2}{\sin(\theta)a_1 + \cos(\theta) a_2}.
\end{equation}

Beam splitters and phase shifters belong to the class of linear optical transformations -- those preserving the total number of bosons ${\bf a}^\dag {\bf a}$. A general linear optics transformation $U$ is represented by
\begin{equation}
\label{unitary}
    \textrm{U}[a_k]= U_{k\ell} \, a_\ell.
\end{equation}
One easily verifies that the canonical commutation relations given before impose unitarity $\textrm{U} \textrm{U}^\dag = \textrm{U}^\dag \textrm{U}=\id$, which is this parametrized by $n^2$ real parameters. The decomposition of any transformation $\textrm{U}$ in beamsplitter and phase shifter can be obtained from the following observation.\\

\begin{figure}[ht!]
	\centering
	\includegraphics[width=0.75\linewidth]{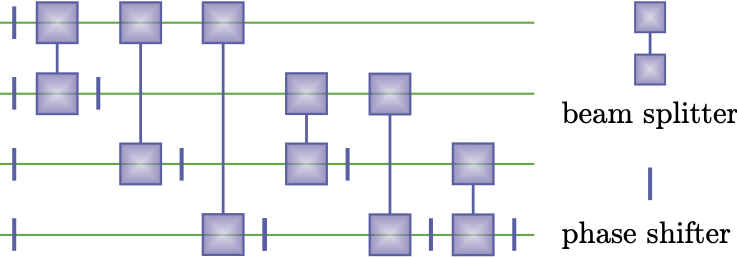}
	\caption{The decomposition of any $n$ mode linear optical transformation $\textrm{U}$ in single-parameter beam splitters and phase shifter. The illustration depicts the case for $n=4$, but one can easily guess the general structure. The mirror symmetric arrangement is also a possible decomposition of any $\textrm{U}$.}
	\label{fig:Udecomposition}
\end{figure}

\textbf{Observation 1} Any linear optical transformation $\textrm{U}$ on $n$ modes can be decomposed in $\frac{n(n-1)}{2}$ beam splitters and $\frac{n(n+1)}{2}$ phase-shifters, as depicted in Fig.~\ref{fig:Udecomposition}. The proof can be deduced from Refs.~\cite{Reck94,He07}. \\

\paragraph{Displacements and squeezing operations---} A displacement -- $\textrm{D}$ -- is a single mode transformation that maps the bosonic operator $a$ into
\begin{equation}
    \textrm{D}_\alpha[a] = a+\alpha,
\end{equation}
where $\alpha= \alpha'+ i \alpha''$ is a complex number. Trivially, we have
\begin{equation}
    \textrm{D}_\alpha = \textrm{D}_{\alpha'}^\textrm{Re} \circ \textrm{D}_{\alpha''}^\textrm{Im} =   \textrm{D}_{\alpha''}^\textrm{Im}\circ \textrm{D}_{\alpha'}^\textrm{Re}.
\end{equation}
The most general combination of displacements on $n$ modes is thus parametrized by $2n$ parameters. It is also easy to see that any displacement can be decomposed as
\begin{equation}
    \textrm{D}_\alpha = \textrm{PS}_{-\varphi}\circ \textrm{D}_{|\alpha|}^\textrm{Re}\circ\textrm{PS}_{\varphi}
\end{equation}
for $\alpha = |\alpha|e^{i \varphi}$.
For compactness we denote a displacement operating on $n$ modes as
\begin{equation}
\label{parallel_D}
    \textbf{D}_{\bm \alpha} = \bigotimes_{i=1}^n \textrm{D}_{\alpha_i}
\end{equation}

Finally, we consider single-mode and two-mode squeezers, the last two \ kinds of Bogolyubov transformations. A single-mode squeezing operation -- SMS -- maps the bosonic operator $a$ onto
\begin{equation}
    \text{SMS}_g[a] = \cosh(g)\, a + \sinh(g) a^\dag.
\end{equation}
For compactness, we introduce the following notation for single mode squeezers operating on $n$ modes
\begin{equation}
\label{parallel_SMS}
    \textbf{SMS}_{\bf g} = \bigotimes_{i=1}^n \textrm{SMS}_{g_i}
\end{equation}

Two-mode squeezing operation -- TMS -- acts on two bosonic modes $a_1$ and $a_2$ as
\begin{equation}
    \textrm{TMS}_g\left[\binom{a_1}{a_2}\right] = \cosh(g) \, \binom{ a_1}{a_2} +
    \sinh(g) \binom{ a_2^\dag}{a_1^\dag}.
\end{equation}

Note that displacement and squeezing operations are not energy preserving, i.e. do not preserve the total number of bosons and do not leave the vacuum state invariant. \\

\paragraph{General Bogolyubov transformation---} It is easy to see that applying previously described elementary operations repeatedly can only map the bosonic operator $a_\ell$ associated to mode $\ell$ to a linear combination of annihilation operators $a_k$, creation operators $a_k^\dag$ and a complex number. The most general transformation $\textrm{G}$ on $n$ modes that has such a form is called a Bogolyubov transformation and can be written as
\begin{equation}\label{eq: G def}
\textrm{G}[a_k]= \sum_\ell (F_{k\ell}\, a_\ell + H_{k\ell}\, a_\ell^\dag) + \alpha_k.
\end{equation}
Unitariry of $\textrm{G}$ ensures that it preserves the commutation relations. It is easy to see that canonical commutation relation enforce
\begin{equation}\label{eq: G cons}
    \begin{split}
     &\text{(i)} \, [a_k,a_\ell^\dag]=\delta_{d\ell}:\qquad\! F F^\dag - H H^\dag =\id\\
     &\text{(ii)}\, [a_k,a_\ell]=0: \qquad F H^T - H F^T =0.
    \end{split}
\end{equation}
These two constraints  leave us with $2 n^2+ 3 n$ parameters. The following observation provides the decomposition of $\textrm{G}$ in terms of basic elements.\\

\textbf{Observation 2}
The general Bogolyubov transformation $\textrm{G}$ on $n$ bosonic modes described in Eq.~\eqref{eq: G def} and satisfying the constraints \eqref{eq: G cons} can be decomposed as 
\begin{equation}
\label{form_G}
     \textrm{G}=\textrm{V} \circ \textbf{SMS}_{\bf g} \circ \textrm{U} \circ \textbf{D}_{\bm \alpha},
\end{equation}
where $\textrm{U}$ and $\textrm{V}$ are linear optics transformations as given in Eq.~\eqref{unitary}, $\textbf{SMS}_{\bf g}$ are parallel single-mode squeezers (see Eq.~\eqref{parallel_SMS}), and $\textbf{D}_{\bm \alpha}$ are parallel displacements (see Eq.~\eqref{parallel_D}). 

Obsevation 2 is at the core of the representation of the covariance matrix representation of Gaussian states and transformations, and relies on the canonical decomposition of symplectic matrices given by Williamson’s theorem, see e.g. \cite{GaussianAdesso, GaussianRMP} for reviews along this line.

A few remark can be made on the decomposition presented in Eq.~\eqref{form_G}. First, we note that any transformation achieved by displacements $\textbf{D}_{\bm \alpha}$ followed by linear optics $\textrm{U}$, can be equivalently represented as $\textrm{U}$ followed by another displacement $\textbf{D}_{\bm \beta}$
\begin{equation}\nonumber
    \textrm{U}\circ \textbf{D}_{\bm \alpha} = \mathbf{D}_{\bm \beta} \circ\textrm{U}.
\end{equation}
Similarly, displacements $\textbf{D}_{\bm \alpha}$ followed by single mode squeezers $\textbf{SMS}_{\bf g}$ are equivalent to the same squeezers followed by other displacements $\mathbf{D}_{\bm \beta}$
\begin{equation}\nonumber
 \textbf{SMS}_{\bf g}\circ \textbf{D}_{\bm \alpha} =  \textbf{D}_{\bm \beta} \circ \textbf{SMS}_{\bf g}.
\end{equation}
Both equivalences follow straightforwardly from transformation of the creation operators $\bf a$.\\

\begin{figure}[ht!]
	\centering
	\includegraphics[width=1\linewidth]{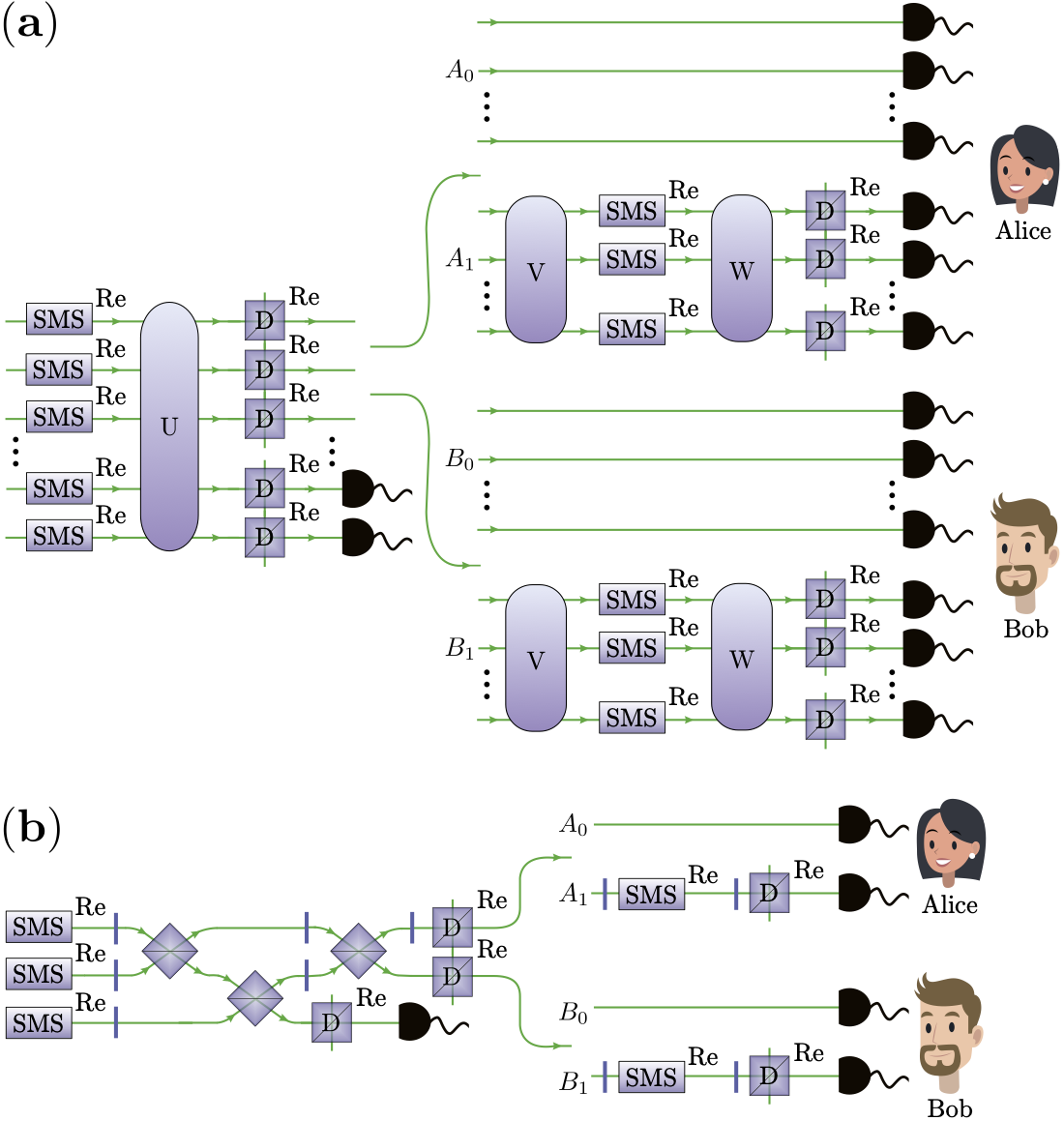}
	\caption{(a) Illustration of the most general setup that is needed to reproduce the Bell inequality violation of any setup in the bipartite case. (b) Illustration of the most general setup able to reproduce any CHSH test in which two parties received one mode each from a preparation stage involving three modes. This shows that even in this fairly simple case, it would take a brute-force optimization over 23 parameters to deduce the setup leading to the highest CHSH score.}
	\label{fig:general}
\end{figure}

We deduce a \textbf{Corollary} of observation 2: Any Bogolyubov transformation $\textrm{G}$ can also be decomposed as
\begin{equation}
    \textrm{G}=\textbf{D}_{\bm \beta}\circ \textrm{V} \circ \textbf{SMS}_{\bf g} \circ \textrm{U}.
\end{equation} 
This decomposition is particularly handy when the transformation $\textrm{G}$ acts on the vaccum $\ket{\bf 0}$, given that linear optics leave the vacuum invariant. We have 
\begin{equation}\label{eq: state decomposition}
    \textrm{G}\ket{\bf 0} = \textbf{D}_{\bm \beta}\circ \textrm{V} \circ \textbf{SMS}_{\bf g} \ket{\bf 0}.
\end{equation}
Any state created by a Bogolyubov transformation from the vacuum can be prepared from single mode squeezed states on which a linear optics transformation and displacement are applied.\\

\paragraph{Complexity of the brute-force search---}
Let us start by considering a setup with fixed measurement settings. From a global perspective, such a setup consists of a state $\textrm{G}\ket{\bf 0}$ prepared by Bogolyubov transformation $\textrm{G}$ acting on vacuum and measured with non-photon number resolving detectors. Before counting the number of parameters that we would need for a brute-force search of the setup leading the highest CHSH score, let us note that the amplitude of the final displacements in the decomposition given in Eq.~\eqref{eq: state decomposition} can be taken to be real.

To see this, note that $\textbf{D}_{\bm \beta} = \textbf{PS}_{-\bm \varphi} \circ\textbf{D}_{|\bm \beta|}^{\textrm{Re}}\circ \textbf{PS}_{\bm \varphi}$, the first raw of phase shifter can be absorbed in $\mathrm{V}$ while the last raw does not affect the measurement statistics. Hence, the most general setup \textit{for fixed settings} is described by the state
\begin{equation}
    \textbf{D}_{|\bm \beta|}^{\textrm{Re}}\circ \textrm{V} \circ \textbf{SMS}_{\bf g} \ket{\bf 0},
\end{equation}
which is specified by $n^2 + 2n$ parameters.

Any additional setting $x$ gives each party $\ell$ (Alice and Bob in the bipartite case), the possibility to apply another local Bogolyubov transformation $\mathrm{G}_x^{(\ell)}$ on the $k_\ell$ received modes. As before this transformation can be decomposed as
\begin{equation}
    \mathrm{G}_x^{(\ell)} = \textbf{D}_{|\bm \beta|}^{\textrm{Re}}\circ \textrm{V} \circ \textbf{SMS}_{\bf g} \circ \mathrm{U},
\end{equation}
which counts $2 k_\ell^2 + 2 k_\ell$ parameters. The total number of parameters required to fully describe an arbitrary setup is thus given by
\begin{equation}
    \#\text{parameters}= n^2 + 2n + \sum_\ell (N_\ell-1)(2k_\ell^2+2k_\ell),
\end{equation}
where $N_\ell$ is the number of settings of the party $\ell$ and $k_\ell$ the number of modes it receives. For the case of the CHSH test, one gets
\begin{equation}
  \#\text{CHSH}= n^2 + 2n + 2(k_A^2+k_A) +  2(k_B^2+k_B).
\end{equation}
The general setup in the CHSH case is illustrated in Fig~\ref{fig:general}(a). Fig~\ref{fig:general}(b) shows in particular the setup that would be needed to reproduce any possible implementation of the CHSH test in the 3 mode case with 2 modes sent to Alice and Bob. This corresponds to one of the examples considered in the main text ($(n,k_A,k_B)=(3,1,1)$). In this case, a brute-force search of the setup leading to the highest CHSH test would require an optimization over 23 real parameters.

\end{document}